\documentclass[reprint,pra,superscriptaddress]{revtex4-1}
\pdfoutput=1
\usepackage{graphicx}    
\usepackage{amsmath}
\usepackage{verbatim} 
\usepackage{tikz}
\usepackage{siunitx} 
\usepackage{titlesec}
\usepackage{mathtools} 
\usepackage{physics}
\usepackage{bm} 
\usepackage{setspace}
\usepackage{hyperref}
\hypersetup{colorlinks,allcolors=blue}
\usepackage[]{siunitx} 
\usepackage[shortcuts]{extdash}
\usepackage{soul}

\begin{document} \title{Magnetic soliton: from two to
three components with SO(3) symmetry} 
\author{Xiao Chai}
\email {xchai@gatech.edu}
\affiliation{School of Physics, Georgia Institute of Technology, 837 State St, Atlanta, Georgia 30332, USA}
\author{Di Lao}
\affiliation{School of Physics, Georgia Institute of Technology, 837 State St, Atlanta, Georgia 30332, USA}
\author{Kazuya Fujimoto}
\affiliation{Institute for Advanced Research, Nagoya University, Nagoya 464-8601, Japan}
\affiliation{Department of Applied Physics, Nagoya University, Nagoya 464-8603, Japan}
\author{Chandra Raman}
\affiliation{School of Physics, Georgia Institute of Technology, 837 State St, Atlanta, Georgia 30332, USA}


\begin{abstract} 
Recent theoretical and experimental research has explored magnetic solitons in binary Bose-Einstein condensates (BECs). Here we demonstrate that such solitons are part of an SO(3) soliton family when embedded within a full three-component spin-1 manifold with spin-rotational symmetry. To showcase this, we have experimentally created a new type of domain wall magnetic soliton (DWMS) obtained by 90 degree rotations, which consist of a boundary between easy-axis and easy-plane polar phases. 
Collisions between SO(3) solitons are investigated by numerically solving the Gross-Pitaevskii equations, which exhibit novel properties including rotation and dissipation of soliton spin polarization.

\end{abstract}

\maketitle
\section{Introduction} 
Bose-Einstein condensates (BECs) of weakly interacting atoms offer a prominent platform for soliton studies \cite{1999:Burger:Lewenstein,2000:Denschlag:Phillips,khaykovich,strecker2002formation}.  In particular, they have become a versatile arena for research on vector solitons, where the rich internal hyperfine level structure of the atoms hosts a great deal of internal symmetry that allows for the observation of many types of solitons in the laboratory \cite{2008:Becker:Sengstock,2001:Busch:Anglin,2018:Bersano:Kevrekidis,nistazakis2008bright}.  Indeed, symmetry has the potential to unify numerous vector soliton observations in the literature \cite{park2000systematic}.  A powerful example of this is the connection between beating dark-dark solitons \cite{2011:Hoefer:Engels,2012:Yan:Cuevas} and dark-bright solitons \cite{2001:Busch:Anglin} which can be connected to one another via SU(2) rotations.  Thus far such symmetry considerations have only been applied to solitons in the integrable, Manakov limit \cite{manakov1974theory}, where the intra-and interspecies interaction strengths are assumed to be equal. A far more diverse range of symmetries is possible in quantum gases with spin dependent interaction, and their connection to vector solitons is a topic that has so far been unexplored.

In this work we uncover theoretically, and provide experimental evidence for a diverse family of solitons in spin-1 BEC with antiferromagnetic interactions.  The recently observed two-component \textit{magnetic} solitons \cite{2020:chai:raman,Farolfi} are shown to be but one member of this family.  We numerically explore the consequences of SO(3) rotational symmetry on this family, as well as its breakdown in the presence of a finite quadratic Zeeman shift.  SO(3) symmetry also creates entirely new possibilities for engineering collisions between solitons of different spin orientation, which we show numerically to have nontrivial spin rotation.  To our knowledge these concepts have only been applied to Manakov systems, although a recent preprint has explored domain wall physics in the case of ferromagnetic interactions  \cite{Yu.Blakie.2020}.

\section{Formalism}
Our work is situated in the mean-field description of a one-dimensional spin-1 BEC relevant to recent experiments \cite{2020:chai:raman,Farolfi}.  It utilizes three coupled Gross-Pitaevskii equations (GPEs),
\begin{align}
	i\hbar \pdv{t} \psi_{m}=\biggl(-\frac{\hbar^2}{2M}\pdv[2]{}{y}&+V \biggr)\psi_m + qm^2\psi_m
	+ g_0 n_{\mathrm{tot}} \psi_m \nonumber \\
  &+ g_2 \sum_{n=-1}^{1} \bm{F} \cdot (\hat{\bm{F}})_{mn}\psi_n,
  \label{eq:gp}
\end{align}
where $\psi_m(y,t)$ is the order parameter with magnetic
quantum number $m=-1,0,+1$. $y,~t$ are space and time
coordinates, respectively. $M$ is the atomic mass.
$V(y),~q$ are the spin independent potential and the
quadratic Zeeman shift, respectively.
$n_{\mathrm{tot}}(y,t)=\sum_{m=-1}^1 \abs{\psi_m(y,t)}^2$
is the total density. $\bm{F}(y,t) =\sum_{m,n=-1}^1
\psi_m^*(y,t)(\hat{\bm{F}})_{mn}\psi_n(y,t)$ is the spin density
and $\hat{\bm{F}}$ is the spin-1 matrix
\cite{spinmatrices}. $g_0,~g_2$ are effective spin
independent and spin dependent interaction coupling
constants in a one-dimensional system, respectively. Here we only consider
repulsive and antiferromagnetic interaction, namely
$g_0>g_2>0$. 

In the absence of external fields, i.e., $V(y)=0,~q=0$, and assuming the total density is a constant $n$,
the above equations have a magnetic soliton solution
\cite{2016:Qu:Stringari,2020:chai:raman,Farolfi} when
$g_2\ \ll g_0$. This condition can be fulfilled in a
sodium BEC where $g_2/g_0 \approx 0.036$
\cite{2011:Knoop:Tiemann}. In a magnetic soliton
solution, the $m=0$ component has no population and the
densities of $m=\pm 1$ are given by
\begin{equation} n_{\pm1} = \frac{n}{2} \left [ 1 \pm
\sqrt{1-U^2} {\rm sech}\left
({\sqrt{1-U^2}}\zeta \right ) \right
], \label{eq:soliton} 
\end{equation}
where $U = V/c_s$ is the soliton velocity
normalized to that of spin waves, $c_s = \sqrt{n g_2/M}$.
$\zeta = (y-Vt)/\xi_s$ is the moving coordinate,
where the width of the soliton depends upon the spin
healing length $\xi_{s} = \hbar/\sqrt{4 M n g_2}$. We can
rewrite this solution in terms of the three-component
spinor,
\begin{equation}
 \bm{\psi}_{\mathrm{MS}} =
\begin{pmatrix}
\sqrt{n_{+1}}e^{i\phi_{+1}} \\ 0 \\ \sqrt{n_{-1}}e^{i\phi_{-1}}
\end{pmatrix},
\label{eq:spinor}
\end{equation}
where the wavefunctions have been written in terms of the
amplitude and phase of each component. In Ref.
\cite{2016:Qu:Stringari}, the phases
$\phi_{A,B} = \phi_{+1} \mp \phi_{-1}$ are shown analytically to be: 
\begin{align}
 \cot{\phi_A}&=-\sinh{(\zeta \sqrt{1-U^2})}/U, \\ \nonumber
\tan{(\phi_B+C)} &= -\sqrt{1-U^2}\tanh{(\zeta \sqrt{1-U^2})}/U,
\end{align}
where $C$ is chosen to ensure $\phi_B(\zeta=-\infty)=0$.
It can be proven that such a solution is asymptotically
valid in the limit $g_2\ll g_0$
\cite{Congy} and we will assume its
deviation from the true solution is negligible.

Under the Cartesian representation \cite{cartesian},
a spin-1 spinor can be decomposed as
$(\psi_x,\psi_y,\psi_z)^t = e^{i\phi}
(\bm{u}+i\bm{v})$, where $\bm{u}$,
$\bm{v}$ are two real vectors obeying
$|\bm{u}|^2+|\bm{v}|^2 = n_{\rm tot}$, while $\phi$ is a
global phase chosen to satisfy $\bm{u}\cdot \bm{v} = 0$
and $|\bm{u}|\ge|\bm{v}|$. Thus a spin-1 state can be
fully characterized by $\bm{u}$ and $\bm{v}$. One can
show that the spin density $\bm{F} = 2\bm{u}\times\bm{v}$
\cite{2016:Zibold:Gerbier,symes2017nematic}, then $\bm{v}
= -(\bm{u}\times\bm{F})/(2|\bm{u}|^2)$, which indicates
that the two vectors $\bm{F},~\bm{u}$ can also determine
a spin-1 state. In a spin-nematic state where $|\bm{F}| =
0$, the vector $\bm{u}$ is called the director whose
direction plays the role of the order parameter
\cite{2016:Zibold:Gerbier}. In this letter, we will
generalize the definition of the director and call
$\bm{u}$ the director in general cases when $|\bm{F}|\neq
0$.

\begin{figure}[htbp]
  \centering
  \includegraphics[width=\linewidth]{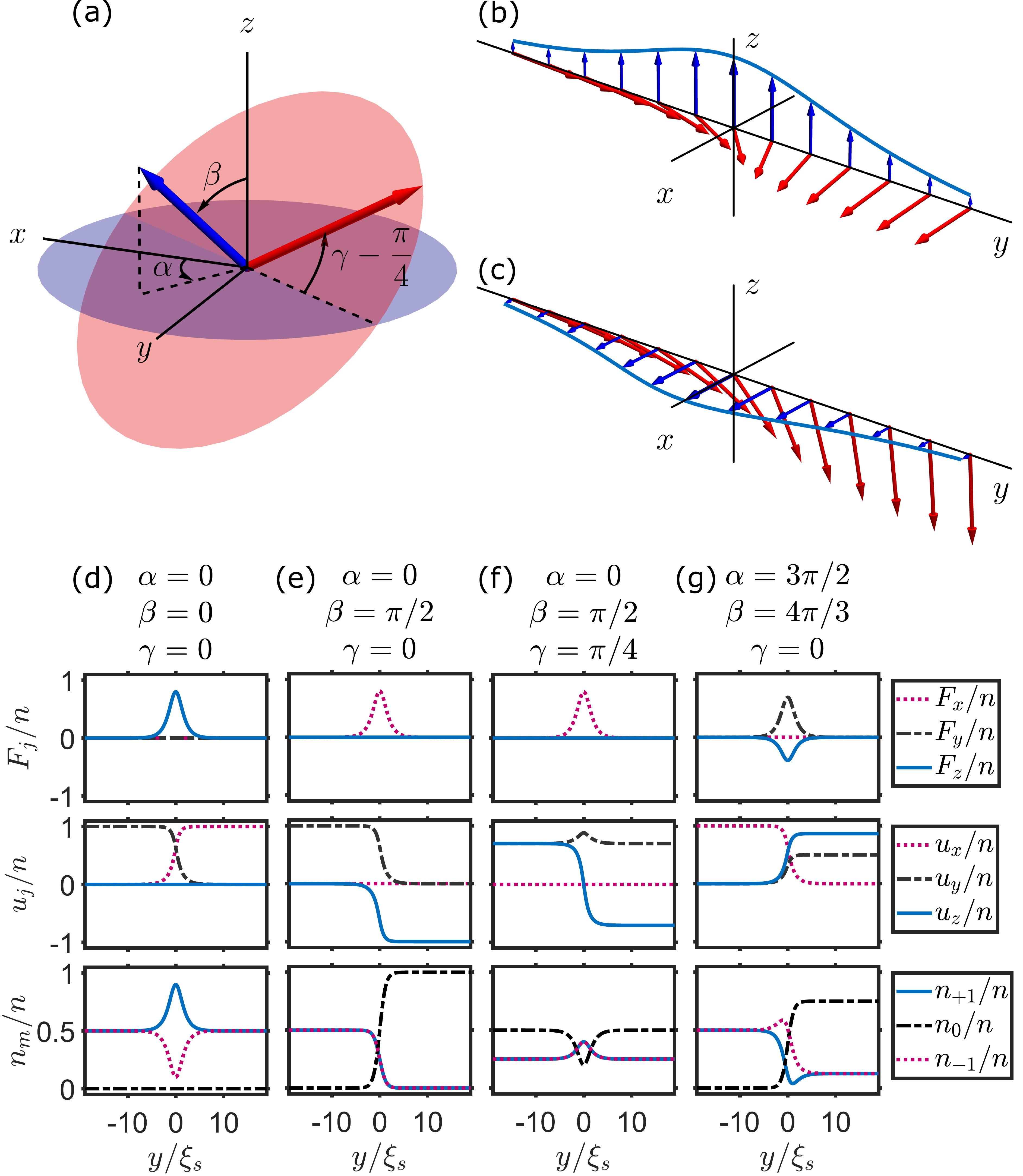}
  \caption{(a) A vector representation of an SO(3)
  soliton. The blue and red arrows are the normalized
  spin density and director at the center of the
  soliton, respectively. The blue disk denotes the
  $x$-$y$ plane. The red disk denotes the plane normal to
  the spin vector, where the director lies on.
  $\alpha,\beta,\gamma$ are Euler angles of the SO(3)
  soliton. The additional angle $\pi/4$ of the red arrow comes from the angle between the $x$-axis and the director before the rotation. (b,c) Spin density field and director 
  field of (b) a magnetic soliton and (c) a domain wall magnetic soliton (DWMS), as described in the main text. The blue and red arrows are the
  spin density and the director, respectively. The
  blue curves are the magnetization profile. (d-g)
  Examples of SO(3) solitons with different Euler angles.
  In all cases the soliton velocity is $U=0.6$. The top
  and middle rows show the spin density and the director,
  respectively. The red dotted, black dashed and blue
  solid lines are vector components in $x,y,z$,
  respectively. The bottom row shows the density
  distributions of the three components, where the red
  dotted, black dashed, and blue solid lines are density
  distributions of $m=-1,0,+1$, respectively.
  }
  \label{fig:example}
\end{figure}

Figure~\ref{fig:example}(b) shows the spatially varying spin density $\bm{F}$ and director $\bm{u}$ for the magnetic soliton solution of Eq.\ (\ref{eq:spinor}) found by Qu et al.\ \cite{2016:Qu:Stringari}.  The spin density points along $\hat{z}$ and is localized near $y=0$, while the director rotates from the $+y$ direction to the $+x$ direction in that same region.  Figure \ref{fig:example}(d) shows a corresponding plot of all 3 components of $\bm{F}$ and $\bm{u}$, as well as the density profiles of the three spin components $n_m,~m=-1,0$ and $+1$.

To extract the essential features of various solitons,
we define the spin vector $\bm{S}$ 
as the normalized spin density at the center of
the soliton,
\begin{equation}
  \bm{S} \equiv \frac{\bm{F}(\zeta = 0)}{n_{\mathrm{tot}}(\zeta = 0)}.
\end{equation}
For the magnetic soliton solution (\ref{eq:spinor}), the
spin vector is given by $\bm{S} = \sqrt{1-U^2}(0,0,1)^t$,
whose amplitude is related to the soliton energy
\cite{2016:Qu:Stringari} as $\epsilon = n\hbar c_s
\abs{\bm{S}}$, while its direction characterizes the 
direction of polarization, which is the $+z$ direction
in this case. 

The spin-1 system described by Eq. (\ref{eq:gp})
conserves the total spin angular momentum and atom number in the absence of external fields, i.e. $q=0$. Thus we
have the symmetry group of the system
\cite{2012:Kawaguchi:Ueda},
\begin{equation}
  G = \mathrm{SO}(3)_{\hat{\bm{F}}} \cross \mathrm{U}(1)_\phi,
\end{equation}
where $\hat{\bm{F}}$ and $\phi$ denote spin and gauge
degrees of freedom. Exploiting the SO(3) rotational
symmetry in the spin degree of freedom, we can construct
a new family of solutions to Eq.~(\ref{eq:gp}) from
$\bm{\psi}_\mathrm{MS}$, namely the \textit{SO(3)
soliton} solutions,
\begin{equation}
  \bm{\psi}_{\mathrm{SO(3)}} = 
  \hat{U}(\alpha,\beta,\gamma)\bm{\psi}_\mathrm{MS},
  \label{eq:so3}
\end{equation}
where $\hat{U}(\alpha,\beta,\gamma)$ is the rotation 
operator acting on the spin degree of freedom and
$\alpha,\beta,\gamma$ are the Euler angles. Eq.~(\ref{eq:so3}) is the principal result of this paper. We will adopt the
$z$-$y$-$z$ convention such that $\hat{U} = 
e^{-i\alpha\hat{F}_z}
e^{-i\beta\hat{F}_y}e^{-i\gamma\hat{F}_z}$. 

Figures~\ref{fig:example}(c,e) show a special case of
the SO(3) soliton when the rotation is about the $y$ axis
for $\pi/2$. This solution has its spin aligned with the $x$-direction.  It no longer has a density bump
(notch) in the $m=+1(-1)$ component, but instead, is a
domain wall separating two sides of the soliton. The director 
rotates from the $+y$ direction to the $-z$ direction across the soliton.
Correspondingly, on one side of $y=0$ is a pure $m=\pm 1$ state asymptotically, while on the other side it is pure $m=0$.  We therefore refer to this particular solution with $x$-polarization as a domain wall magnetic soliton (DWMS).

The SO(3) soliton solution (\ref{eq:so3}) is parametrized
by four free parameters $\alpha,\beta,\gamma,U$. In Fig.~\ref{fig:example}(a) we show the spin vector and the director at the center of an SO(3) soliton. Notice
that after an SO(3) rotation, the spin vector is
given by $\bm{S} = \sqrt{1-U^2}(\sin{\beta}\cos{\alpha},
\sin{\beta}\sin{\alpha}, \cos{\beta})^t$, that is to say,
the first two Euler angles $\alpha, \beta$ control the
azimuthal and polar angles of the spin vector of an SO(3)
soliton, as illustrated in Figs.~\ref{fig:example}(d,e,g).
It also indicates that the soliton amplitude is still
related to its velocity as $\abs{\bm{S}} = \sqrt{1-U^2}$. The energy of an SO(3) soliton is still $n\hbar c_s \abs{\bm{S}}$ thanks to the invariance of energy under SO(3) rotations.
Comparing Fig.~\ref{fig:example}(e) and
Fig.~\ref{fig:example}(f), one can find that the Euler
angle $\gamma$ has no effect on the spin, but modifies
the density distribution and the director of a soliton.

\section{Experiment}
\begin{figure}[htbp]
  \centering
  \includegraphics[width=\linewidth]{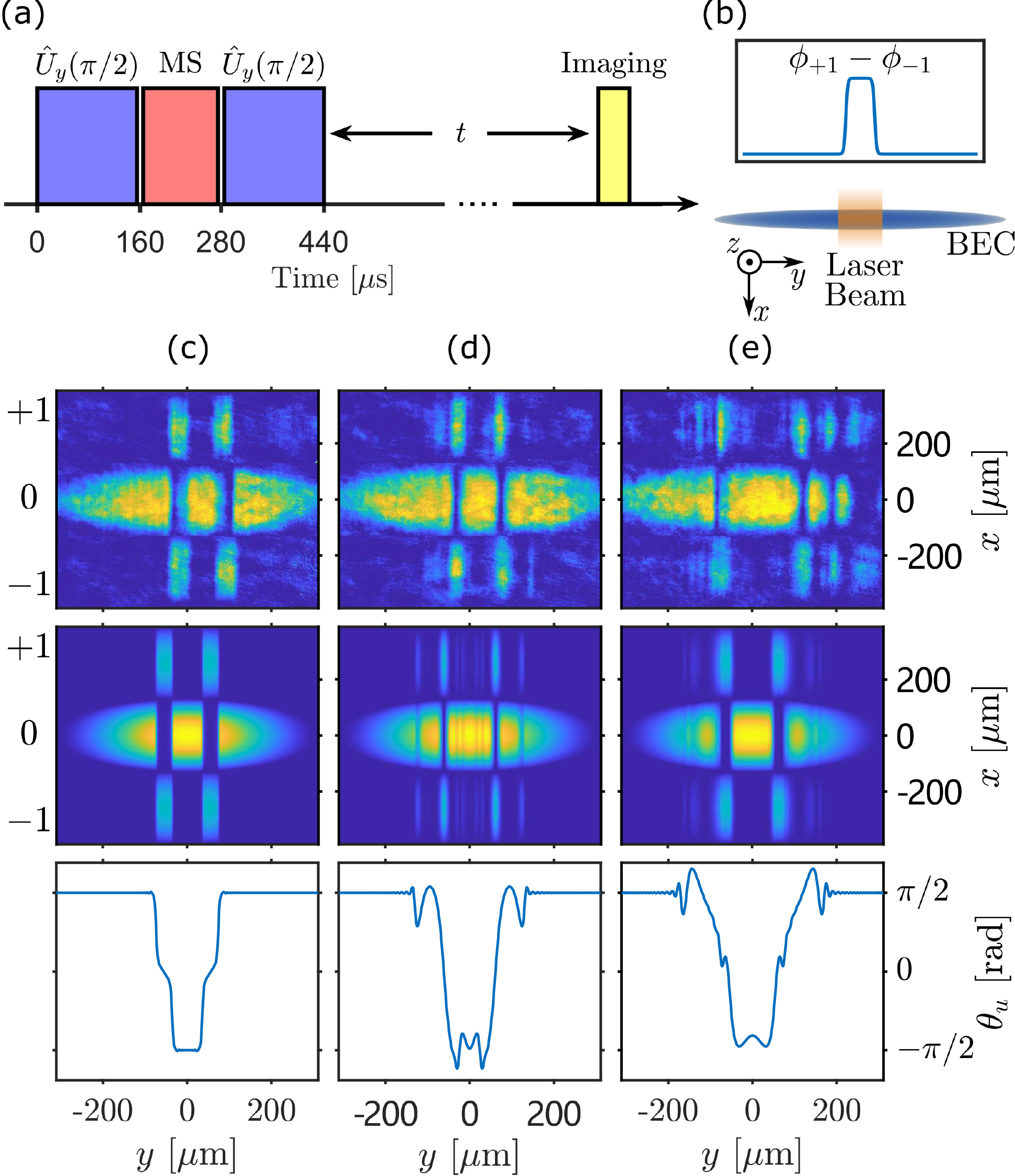}
  \caption{(a) Experimental sequence for generation of
  DWMSs in spin-1 BECs. Initially the condensate
  is prepared in the $m=0$ state. $\hat{U}_y(\pi/2)$
  represents a spin rotation about $y$ for $\pi/2$. MS
  means "magnetic shadow". (b) Magnetic shadow. A
  circularly polarized and properly detuned laser beam
  illuminates the $m = \pm 1$ BEC along the $z$
  direction for a short duration. The pulse induces
  different Larmor precession rates of the two
  components, leading to local relative phase gradients.
  The laser beam is masked by two knife edges, resulting
  in a magnetic shadow on the condensate. (c-e) Evolution of DWMSs in a harmonic trap. The evolution times are 
  (a) \SI{15}{\milli\second}, (b)
  \SI{50}{\milli\second} and (c) \SI{90}{\milli\second}. The
  top panel shows Stern-Gerlach time-of-flight images of the condensates
  after the generation of DWMSs. The middle panel shows corresponding 1D 
  simulation results with $q=\SI{1.43}{\hertz}$, where the
  time-of-flight effects are added manually for visual
  aid. The simulation time is slightly different from the experimental hold time for reasons discussed in the main text. The bottom panel shows the angle $\theta_u$ between
  the director and the $y$ axis. In the simulation the diretor is always on the $y$-$z$ plane.} \label{fig:exp}
\end{figure}

Here we report experimental evidence for the domain wall magnetic soliton (DWMS), one member of the SO(3) soliton family.
The experiments
began with a sodium BEC of $5 \times 10^6$ atoms in an
elongated dipole trap with oscillation frequencies of
$(\omega_x,\omega_y,\omega_z) = 2\pi \times
(380,5.4,380)~\si{\hertz}$. A bias magnetic field of
$72\pm4~\mathrm{mG}$ was applied along the $z$ direction
which defined the quantization axis. Initially all atoms
were prepared in the $m=0$ state. To generate DWMSs, we first applied an RF pulse along the $y$
direction (see Fig.~\ref{fig:exp}(a)), which effectively
rotated the spin of the BEC by 90 degrees, resulting in
an $m=\pm 1$ mixture. Subsequently, we cast a magnetic shadow using a far-off-resonance laser beam with a top-hat beam profile as shown schematically in Fig.~\ref{fig:exp}(b).  As discussed in \cite{2020:chai:raman}, each edge of the top hat generated a pair of 
$\pm z$-oriented SO(3) solitons propagating in opposite
directions. To create the DWMSs, we instead applied another RF rotation
about $y$ by 90 degrees immediately after the magnetic shadow sequence.  This realizes DWMSs (SO(3) solitons with $\alpha=0$ and $\beta=\pm \pi/2$), which evolve in the trap for a variable hold time before detection. This method is successful if the system retains spin coherence between the two RF pulses.  Experimentally we have found this to be true for pulse separations up to \SI{200}{\micro\second}, and limited by ambient magnetic field fluctuations at the mG level.

Figures~\ref{fig:exp}(c-e) display Stern-Gerlach images of
the condensates for different hold times. Two local
relative phase gradients are imprinted in the
condensates, as shown in Fig.~\ref{fig:exp}(b).
Consequently, we expect four domain walls to be generated,
exactly as one sees in the density distributions in the images taken for 
$\SI{15}{\milli\second}$ and $\SI{50}{\milli\second}$
hold times. The image with $\SI{90}{\milli\second}$ hold
time displays a more complicated pattern.  We believe this is primarily due to the residual quadratic Zeeman shift of $q = 1.43$ Hz present in the experiment.
A nonzero value of $q$ has the effect of reducing the symmetry of the spinor condensate from SO(3) to SO(2) \cite{2012:Kawaguchi:Ueda}. Consequently, the DWMSs
are no longer stationary states.

As in our earlier work \cite{2020:chai:raman}, we compared our results to numerical simulations of the spin-1 Gross-Pitaevskii equations.  We numerically imprinted a double-edged laser beam on the $m=\pm1$ BEC, and rotated it by $\pi/2$ along $y$-axis with an instantaneous pulse.  The system evolved for 100 ms in the presence of a $q = 1.43$ Hz quadratic Zeeman shift, such that $q/g_2n_{\rm tot}(0)\simeq 0.012$. We find from simulation results in Fig.~\ref{fig:exp}(c,d) that the angle $\theta_u$ between the director and $y$-axis shows a $\pi/2$ rotation across the generated domain wall structures, which is the characteristic feature of an SO(3) soliton. The angle of the director in Fig.~\ref{fig:exp}(e) shows no sharp jump for the outer two DWMSs, indicating the decay of those two solitons. As shown in the second row of Figs.~\ref{fig:exp}(c-e), the simulation matches the experiments well qualitatively  with residual differences attributed to inhomogeneity of the phase imprinting beam and its power, and imperfect RF rotation.
\section{Stability}
Contrary to our initial expectation that the soliton pair should decay under finite quadratic Zeeman energy $q$, the experimentally observed DWMSs show a quasi-stable nature. A simple qualitative argument helps us to understand the role of $q$ in the ensuing dynamics.  As the DWMSs created by each edge of the laser beam propagate away from each other, they expand the $m=\pm1$ component.  For $q>0$, as in the experiment, this increases the system energy at the expense of soliton kinetic energy.  Thus the quadratic Zeeman shift acts as a ``trap'' for the solitons.  For small $q$ the trap is insufficient to contain the solitons but slows them down, so that they eventually decay into multiple solitonic structures as shown in the experimental data and the simulations.  For large $q$, e.g. $q=0.3g_2 n$, by contrast, the stability can be {\em enhanced}--we have numerically observed trapping of two DWMSs to form very stable structures.

\begin{figure}[htbp]
  \centering
  \includegraphics[width=\linewidth]{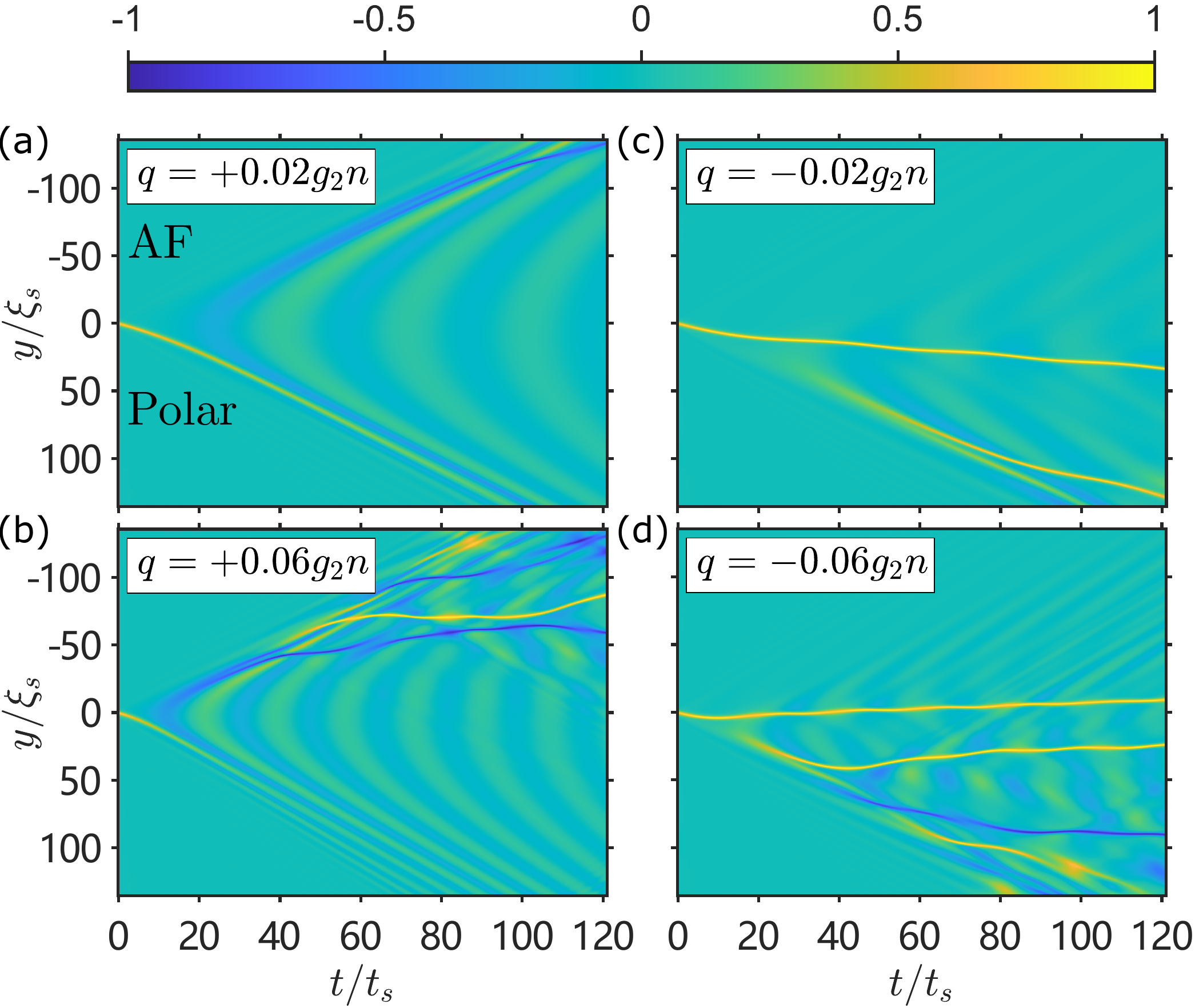}
  \caption{Space-time plot of $F_x/n$ of the SO(3) soliton by rotating $\pi/2$ around \textit{y} axis with different quadratic Zeeman energy (a) $q/g_2n=0.02$, (b) $q/g_2n=0.06$, (c) $q/g_2n=-0.02$ and (d) $q/g_2n=-0.06$. Here the time scale is $t_s=\xi_s/c_s$. The initial condition is a DWMS given in Fig.~\ref{fig:example}(e). It shows that there are more extra magnetic structures besides the DWMS appearing in (b) and (d) than (a) and (c), which implies that the rotated soliton decays faster as the magnitude of quadratic Zeeman energy $|q|$ increases.The emergence of more magnetic structures at $y<0$ (occupied by $m=\pm1$ atoms) in (b) and at $y>0$ (occupied by $m=0$ atoms) in (d) indicates that (b) the antiferromagnetic and (d) the polar part of the DWMS is unstable.
  }
  \label{fig:mag_stab}
\end{figure}

Our data clearly show that the pairs of DWMSs can be quite stable in the trap, propagating freely for timescales of 50 ms or so. Their structure at longer times is clearly influenced by the nonzero value of the quadratic Zeeman shift present in the experiment. This is to be expected, since the Hamiltonian is no longer SO(3)-invariant. To understand this effect in greater detail we performed numerical simulations of single soliton dynamics. These highlight the role of spin-exchange collisions on the soliton stability.  Fig.~\ref{fig:mag_stab}(a-d) shows the magnetization profile of single DWMS evolving with different $q$.  The initial state is a single magnetic soliton in a homogeneous BEC with coherent superposition of $m=\pm1$ that has been rotated by $\pi/2$ along \textit{y} axis. Multiple magnetic structures show up during the evolution of the soliton, indicating the soliton is unstable and decaying. As the magnitude of quadratic Zeeman energy $|q|$ increases, an increasing number of magnetic structures appear and the DWMS becomes more unstable. The quadratic Zeeman energy is set up to $|q|/g_2n=1$ in our simulation, the dynamics of the DWMS is similar to the one in small $q$ regime shown in the Fig.~\ref{fig:mag_stab}. The density profile of this DWMS is shown in Fig.~\ref{fig:example}(e), which can be naively understood as the superposition of a polar part ($m=0$ atoms) occupying $y>0$ region and antiferromagnetic part ($m=\pm1$ atoms) occupying $y<0$ region, with a small overlap around $y=0$. When $q>0$, the magnetic structures emerge mainly at $y<0$ region of the condensate occupying by the $m=\pm1$ atoms since the spin-exchange collisions $|0\rangle+|0\rangle \rightarrow |+1\rangle+|-1\rangle$ in the polar part is suppressed by the energy barrier $q$. While in the $q<0$ regime, more magnetic structures are populated in the $m=0$ region ($y>0$) since the spin-exchange collisions are energetically favored in the polar part of the soliton.

\begin{figure}[htbp]
  \centering
  \includegraphics[width=\linewidth]{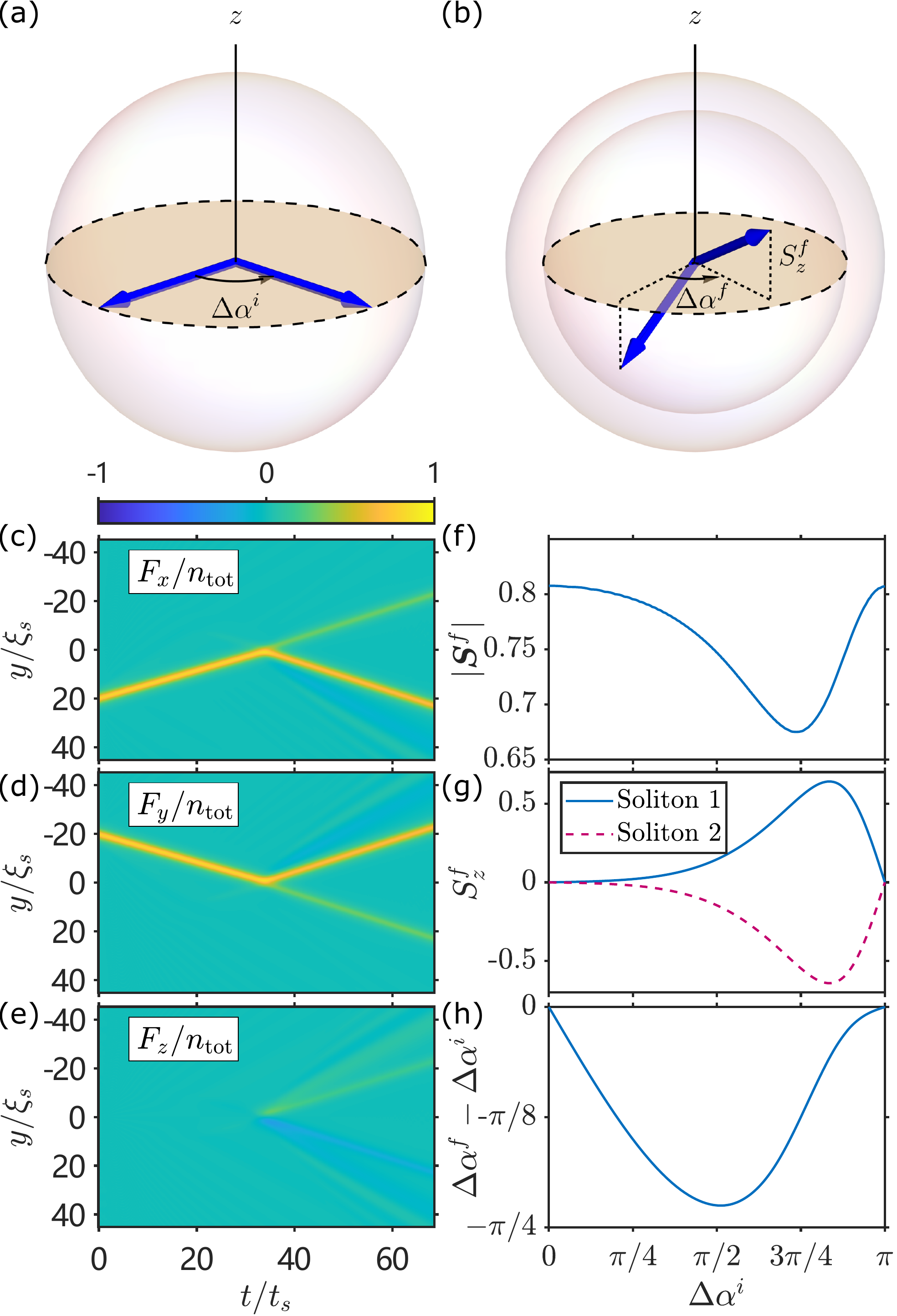}
  \caption{Collisions of SO(3) solitons. (a) Initially 
  the spin vectors represented as two arrows are on the $x$-$y$ plane (the brown disk) with an angle
  $\Delta \alpha^i$ between them. (b) After the
  collision, the spin vectors of the two solitons precess
  out of the $x$-$y$ plane, while both the azimuthal
  angle difference $\Delta \alpha^f$ and the soliton
  amplitudes $|\bm{S}^f|$ decrease. (c-e)  Collisions between two SO(3) solitons whose parameters are
  $(\alpha=0,\beta=\pi/2,\gamma=0,U=-0.6)$ and
  $(\alpha=\pi/2,\beta=\pi/2,\gamma=0,U=0.6)$,
  respectively. Plotted are normalized spin densities in
  the three directions as a function of space and time.
  Here the time scale is $t_s = \xi_s / c_s$. (f)
  Dependence of soliton amplitudes after collisions
  on the initial angular difference $\Delta \alpha^i$.
  (g) The blue solid and red dashed lines are $z$
  components of the spin vectors of the two solitons
  after collisions, respectively. (h) Change of $\Delta
  \alpha$ as a function of $\Delta \alpha^i$.}
  \label{fig:collision}
\end{figure}
\section{Collision}
Finally, we return to the study of soliton collisions at $q = 0$, where striking properties emerge. We have observed that the collisions, while only slightly elastic in nature, cause a nontrivial rotation of the direction of the spin vectors.
We numerically examine these collisions in a uniform  
system. Initially 
two SO(3) solitons are prepared in the condensate,
separated by $40 \xi_s$ and moving towards each other
with $U = \pm0.6$ (so that the initial amplitudes of the two solitons are the same, $\abs{\bm{S}^i}=0.8$). The spin vectors of the two solitons 
are on the $x$-$y$ plane with an angle $\Delta \alpha^i$
between them, as illustrated in
Fig.~\ref{fig:collision}(a). After the collision, as
shown in Fig.~\ref{fig:collision}(b), the spin vectors
precess out of the $x$-$y$ plane and also nod towards 
each other. As an example, Figs.~\ref{fig:collision}(c-e)
show typical collisional dynamics of two SO(3) solitons
with spin vectors initially along $+x$ and $+y$,
respectively. Spin waves are radiated during the
scattering of the solitons, indicating the inelastic
nature of such collisions. 


We then examine the dependence of SO(3) soliton
collisions on the initial angle $\Delta \alpha^i$ between
the spin vectors of the two solitons.
Figure~\ref{fig:collision}(f)
illustrates the amplitude of the spin vectors after a collision as a function
of $\Delta \alpha^i$. Since the energy of an SO(3) soliton is proportional to its spin vector amplitude $\abs{\bm{S}}$, decrease of $\abs{\bm{S}}$ indicates energy loss. Although the collisions with
$\Delta \alpha^i = 0$ or $\pi$ are elastic, in general
cases the solitons have energy loss after a collision,
and the loss is maximized at a certain angle. We suspect the inelastic nature of SO(3) soliton collisions is due to the non-integrability of our system.
Figure~\ref{fig:collision}(g) shows the collisions induce
a non-zero $z$ component of the spin vectors. Because of 
the conservation of total spin, the $z$ components are 
opposite for the two solitons after the collision. The
two solitons undergo nutation as well, where the nutation angle is shown in
Fig.~\ref{fig:collision}(h). We remark that the polarization shift of SO(3) solitons induced by collisions is quite similar to that of dark-bright-bright
solitons in a three-component Manakov system
\cite{prinari2015dark,2020:Lannig:Oberthaler}. Spin precession is also reported for solitons in an integrable spin-1 system \cite{2004:Ieda:Wadati}. Such similarities indicate possible universal connections between different types of solitons carrying spin. Unlike our system, the integrability of a Manakov system or an integrable spin-1 system guarantees the conservation of energy of solitons after collisions. 

\section{Conclusion}
In summary, we have constructed, and experimentally observed signatures of a family of SO(3) soliton solutions to the three-component spin-1 GPEs for antiferromagnetic interactions. 
We believe our work will invoke follow-up experimental 
studies of SO(3) soliton collisions. Theoretical problems, e.g. the physical interpretation of such collisional behaviors in non-integrable systems remain to be explored.  The stability of SO(3) solitons created under controlled conditions in higher spatial dimensions is an intriguing avenue for future exploration.


\section{Acknowledgement}
This work acknowledges NSF Grant No. 1707654, and JSPS KAKENHI (Grants No. JP19H 01824, No. JP19K14628, No. 20H01843), and the Program for Fostering Researchers for the Next Generation (IAR, Nagoya University) and Building of Consortia for the Development of Human Resources in Science and Technology (MEXT).

%

\end{document}